# Factors controlling protein evolvability—at the molecular scale

Jorge A. Vila

IMASL-CONICET, Universidad Nacional de San Luis, Ejército de Los Andes 950, 5700 San Luis, Argentina.

## Abstract

This piece serves two purposes. Firstly, it aims at elucidating the role of 'epistasis' in shaping—at a molecular level—the evolutionary paths of proteins, as well as the extent to which these epistatic effects are the outcome of an as-yet-unidentified epistatic force. Second, it seeks to ascertain the extent to which the 'principle of least action' will enable us to identify which of all potential trajectories has the highest evolutionary efficiency, as well as how variations in factors such as protein robustness and folding rates—resulting from the unavoidability of destabilizing mutations—might influence this critical evolutionary process. The initial findings suggest that protein evolution—at a molecular level—may be more predictable than previously thought, as 'epistasis' and the 'principle of least action' collectively impose constraints on evolutionary paths and trajectories, and consequently, on protein evolvability. Thus, this work should advance our understanding of the main molecular mechanisms that underlie the evolution of mutation-driven proteins and also provide grounds to answer a fundamental evolutionary question: how does Darwinian selection regard all potential trajectories available?

**Introduction**

Ever since its inception (Bateson, 1909), the term 'epistasis'—which is usually understood as how the genetic background affects mutations or how the combined impact of multiple mutations differs from simply adding their individual effects—has received substantial attention in the literature throughout the years (Vila, 2024a, and references therein). Additionally, the effects of their epistatic interactions have been classified by using an assortment of names (see Domingo *et al*., 2019), such as being 'direct or indirect' (Lipsh-Sokolik & Fleishman, 2024), 'positive or negative' (Bonhoeffer *et al*., 2004; Miton & Tokuriki, 2016; Domingo *et al*., 2019), 'specific or nonspecific' (Starr & Thornton, 2016), or having the potential to result in sign epistasis (a change in the direction of epistatic effects from positive to negative and vice versa; Weinreich *et al*., 2005; Zhang *et al*., 2024). All of the above, combined with the growing evidence from fitness landscape theory (Bank, 2022) that epistasis may affect the path and outcome of evolution (Sailer & Harms, 2017), lead us to question whether these effects are all the result of an epistatic force that has not yet been identified. The possible existence of such a 'force' should be discussed under the light of mounting evidence that the majority of the effects mentioned attributed to epistasis can be concisely explained by acknowledging how they depend on two crucial variables: the milieu and amino acid sequence, which ultimately prove to be the main factors of the protein folding problem (Vila, 2024a). Clarification of this query will be crucial, among other things, to evaluate the role of epistasis in determining evolutionary trajectories, which is one of the main purposes of our analysis. Complementary to the above, we also plan to assess whether, and how, the physical principle of least action—a basic variational principle of particle and continuum systems (Feynman *et al*., 1963; Landau & Lifshitz, 1975; Hanc & Taylor, 2004)—could shape protein evolutionary trajectories. The analysis will help us evaluate how changes due to mutations in both protein folding rates (Vila, 2023b) and protein robustness (Vila, 2024b) affect predictions of protein trajectories at the molecular level. Additionally, the analysis will provide us with an opportunity to discuss a long-standing question in evolutionary biology: What is the role of robustness in protein evolvability? (Taverna & Goldstein, 2002; Wagner, 2005a; Wagner, 2005b; Bershtein *et al*., 2006; Bloom *et al*., 2006; Tokuriki & Tawfik, 2009; Bloom & Arnold, 2009; Arnold, 2009; Romero & Arnold, 2009; Wagner, 2008; Masel & Trotter, 2010; Tóth-Petróczy & Tawfik, 2014; Mayer & Hansen, 2017).

To achieve the aforementioned objectives, we should start by providing some agreement on the nomenclature used, because it will be crucial for a clear understanding of the process of protein evolution at the molecular level. A notable example of the latter is the differentiation between 'trajectories' and 'paths,' which are not interchangeable terms. A path is a spatial construction made up of a series of steps, each of which represents a single-point mutation, but it lacks temporal information (see Figure 1**a**). A trajectory, in contrast, is a path defined by temporal data that specifies the speed—at each step—and duration of movement along it (see Figure 1**b**). Concerning the latter, it is worth noting that we chose a particular β-lactamase as our model protein for the present study—which aims to investigate the factors shaping protein evolutionary paths and trajectories—because it increases bacterial resistance by a factor of ~100,000 against a clinically important antibiotic such as cefotaxime (Weinreich *et al.*, 2006).

Despite the abundance of literature on the definitions of 'evolvability' and 'robustness' (Wagner, 2008; Masel & Trotter, 2010; Tóth-Petróczy & Tawfik, 2014; Mayer & Hansen, 2017), we have chosen the following ones here: Evolvability describes the ability of a protein to change both its sequence and function over time, whereas robustness indicates the natural capability of a protein to survive single-point mutations without losing functionality—something that will happen if the marginal stability change upon mutation of the protein goes beyond a universal threshold of $|\Delta\Delta G| \sim 7.4$ kcal/mol, a point beyond which a protein may unfold or become non-functional (Vila, 2021, 2022). It should be noted that another definition of robustness includes the ability of proteins to tolerate mutations while maintaining the original structure and function (Tóth-Petróczy & Tawfik, 2014). We opted to exclude the requirement of "…while maintaining the original structure…" from our definition of robustness, since any mutations will result in structural alterations in the native state of the protein—though not necessarily affecting its function—as demonstrated by the amide hydrogen-exchange protection factors (Hvidt & Linderstrøm-Lang, 1954; Privalov & Tsalkova, 1979; Craig *et al.*, 2011; Englander, 2023), which act as a highly sensitive indicator for their detection (Vila, 2022). Overall, there are numerous alternate definitions for robustness and evolvability, as well as disagreements over whether they are positively or negatively correlated (for more information on this topic, see Wagner, 2008 and Mayer & Hansen, 2017).

Another crucial aspect to emphasize from the start is the motivation for the current work. We had shown (Vila, 2024b) that for a convergence protein evolution model—when all paths result in the same target sequence (Ogbunugafor, 2020), such as for two β-lactamase variants separated by five mutations (Weinreich *et al.*, 2006)—the native-state-protein-marginal stability for both the starting wild-type sequence ($\Delta G_{wt}$) and the end target sequence ($\Delta G_{ts}$) is the only information that is needed to analyze the evolutionary process because the total change in free energy ($\Delta\Delta G$) is a state function. From this thermodynamic perspective (Vila, 2025) distinguishing between paths and trajectories and considering epistasis effects is unnecessary (see Figure 2**a**). This viewpoint is invaluable if we focus only on examining evolutionary outcomes. However, such an analysis is useless if our interest focuses on determining the underpinning factors ruling protein evolution. Because one of our main goals is to identify the key factors governing the most likely protein evolution paths and trajectories—as this could be vital to understanding critical biological mechanisms at the molecular level—we decided to take a different approach (see Figure 2**b**). To put it another way, we decided to seek why nature prefers one evolutionary trajectory over another. The solution for this will help us better understand how proteins evolve and, more importantly, will allow us to modify them for specific purposes, which is the aim of directed evolution applications (Romero & Arnold, 2009; Arnold, 2009).

It is important to recognize that evolutionary trajectories are influenced not only by molecular constraints, as will be examined here, but also by biological frameworks (Poelwijk *et al.*, 2007; Jubb *et al.*, 2017; Dishman & Volkman, 2018; Vila, 2020; Buda *et al.*, 2023b; Di Bari *et al.*, 2024; Fuentes-Ugarte *et al.*, 2025). However, a comprehensive analysis of these factors exceeds the scope of the present study.

**Does any specific force trigger epistatic interactions?**

Let us be clear from the beginning: there is no single force associated with epistatic interactions. In contrast to the widely acknowledged interactions resulting from natural forces such as Coulombic, van der Waals and gravitational forces (Einstein & Infeld, 1961; Israelachvili, 1985), epistatic interactions are the result of all forces at play right after either a single-point mutation or environmental changes, a phenomenon that requires examining them as an 'analytic whole' rather than as a mere aggregation of components (Vila, 2024a). Hence, the epistasis origin precludes identifying any particular 'force' as responsible for the epistasis effect. What is the relevance of this conclusion? The answer to this inquiry is straightforward: "*... understanding the forces that shape protein evolution has been a longstanding goal of evolutionary biology*" (Bloom

& Arnold, 2009). Therefore, it is highly important to acknowledge that only after determining how the amino acid sequence encodes their folding—the solution to the protein folding problem—can a solution to any epistatic interaction be accurately determined (Vila, 2024a). So, every mutation (except for a silent one) will cause epistatic interactions, no matter if the epistasis effects are 'direct or indirect,' 'positive or negative,' 'specific or nonspecific,' 'strong or weak.' This is easy to understand: the forces that hold the folded protein will be rearranged even if the protein sequence changes by just one amino acid (Pauling *et al.*, 1949; Ingram, 1957; Eaton, 2020; Shortle, 2009). This observation supports our conjecture about the genesis of 'epistatic interactions' (Vila, 2024a). As a result, we are still unable to accurately forecast epistatic interactions based on mutations. Notwithstanding this limitation, we can predict the epistatic effects that mutations will generate. This is feasible due to the presence of very sensitive indicators of structural changes, such as variations in observable metrics like protein folding rates (Vila, 2023b) or amide hydrogen-exchange protection factors (Vila, 2022). Indeed, analyzing a large-scale experiment on single-point mutations (Tsuboyama *et al.*, 2023) allowed us to demonstrate the sensitivity of such parameters to the resulting slight free energy change ($\Delta\Delta G$) on protein marginal stability (Vila, 2024b). A similar conclusion will be drawn if there are no mutations but there are changes in the environment (milieu), which may cause modifications in both structure and function, as seen in metamorphic proteins (Vila, 2020).

Overall, the genesis of epistatic interactions can only be traced to all of the forces acting during the protein folding process, not to any individual one. The rationale for the latter can be found in Anfinsen's dogma or thermodynamic hypothesis, which states that a protein native state is "*the one in which the Gibbs free energy of the whole system is lowest*" (Anfinsen, 1973)—a concept that defines the physics underpinning protein evolution at the molecular level (Vila, 2025). A proper understanding of this issue will allow us to examine—without bias—the influence and impact of the factors governing protein evolvability, as well as their involvement in determining the most probable evolutionary paths, as will be discussed in the next section. Before moving forward, it is important to note that the use of concepts such as epistasis, and epistatic interactions' (and the numerous associated effects)—as practical tools for protein function design (Lipsh-Sokolik & Fleishman, 2024) or evolutionary process analysis (Phillips, 2008; Zhang *et al.*, 2024)—is not dismissed or unadvised as a result of the analysis above.

### Factors that outline the most likely paths and trajectories

Here, we are going to examine the role and impact of both the least action principle and epistasis on determining the most likely paths and trajectories of an evolutionary process—such as the β-lactamase problem (Weinreich et al., 2006).

#### *The role of the least action principle*

The investigation of the evolutionary paths connecting two natural β-lactamase variants (Weinreich *et al*., 2006) reveals that the majority of the five intermediate mutations on each successful evolutionary path encode destabilizing mutations that are, nevertheless, critical to the functional transition (Lipsh-Sokolik & Fleishman, 2024). The latter observation should not be surprising since most mutations that confer new functions have been proven to be predominantly destabilizing (Condra *et al*., 1995; Blance *et al*., 2000; Wang et al., 2002; Bloom *et al*., 2006; Tokuriki *et al*., 2008; Jensen *et al*., 2018; Domingo *et al*., 2019). Given that destabilizing mutations accelerate the folding/unfolding process (Vila, 2023b), the following questions arise: What impact would the protein folding rate change—resulting from mutations—have on determining the most likely trajectory? Does this issue have any bearing on the 'principle of least action'? Let us start by introducing the physical principle and then give answers to these crucial questions. The principle of least action in physics (Feynman *et al*., 1963; Landau & Lifshitz, 1975; Hanc & Taylor, 2004) states that a system will naturally follow the path that minimizes the 'action' (a functional of the trajectory) among all possible trajectories between two given points. Then, the least action principle enables us to determine a trajectory for which the variation ($\delta$) of the functional ($\delta\Gamma$) is zero, with $\Gamma$ representing the functional of the trajectories. In other words, the system is predicted to evolve in the direction that minimizes the functional. The principle essentially suggests that nature operates with maximum 'efficiency.' This raises a question: how does a possible trajectory of a convergent evolutionary problem become the most probable—and efficient—one? More precisely, how does it apply to the β-lactamase problem? Before we discuss this particular case—constrained to have 5 mutational steps connecting one allele of the wild-type β-lactamase gene (TEM$^{wt}$; Ruiz, 2018) to the target sequence (TEM$^*$) via 10 most-probable trajectories (see Figure 2 of Weinreich *et al*., 2006)—we will discuss a general evolutionary problem. For example, for a convergent evolutionary process (Vila, 2025) consistent of an arbitrary number of mutation (*j*) the functional ($\Gamma_j^\mu$) can be defined for each evolutionary trajectory ($\mu$) in terms of the folding rate $\tau_k$ (as shown in Fig. 1b), as:

$$\Gamma_j^\mu = \sum_{k=1}^{j} \tau_k^\mu = \tau_{wt} \sum_{k=1}^{j} (\tau_k^\mu / \tau_{wt}) \sim \tau_{wt} \sum_{k=1}^{j} e^{\beta \Delta \Delta G_k^\mu} \quad \text{(sec)} \qquad (1)$$

where $\tau_k^\mu \sim \tau_0 \, e^{\beta \Delta \Delta G_k^\mu}$, represent the time to overcome the free-energy barrier $\Delta G_k^\mu$ (Vila, 2023b); $\tau_0$ a pre-exponential factor (Vila, 2023b); $\beta = 1/RT$, $R$ stands for the gas constant and $T$ for the absolute temperature in Kelvin degrees; $\Delta \Delta G_k^\mu = (\Delta G_k^\mu - \Delta G_{wt})$ the free-energy change after the $k$-mutational step respect to that of the wild-type ($wt$) protein (Vila, 2023b); and $\Delta G_j = \Delta G_{ts}$, is the free-energy barrier of the target sequence ($ts$). Let us consider two final thoughts about Equation (1). Firstly, the elapsed time between mutations is assumed to be the same for all possible trajectories. Secondly, from the perspective of the least action principle, the most efficient (and probable) trajectory will be one for which the functional ($\Gamma_j^\mu$) satisfies the following condition: $\delta \Gamma_j^\mu = 0$. In other words, the stationary point may be a minimum or a saddle point, but not a maximum. Therefore, the most efficient trajectory should have as many destabilizing mutations as possible, as each will diminish the contribution to Eq. (1)—given that $\Delta \Delta G_j^\mu < 0$—and thereby minimize the functional ($\Gamma_j^\mu$). This suggestion is supported by evidence (Wang *et al*., 2002) regarding β-lactamase intermediate 5-step mutations, indicating that certain mutations—such as G238S and E104K—result in a loss of both thermodynamic stability and kinetic activity compared to the wild-type enzyme (TEM$^{wt}$), whereas other mutations—like M182T—do not affect activity but restore thermodynamic stability. These results refer to three of the five mutations—A42G, E104K, M182T, g238s, and G4205A—shared by the 10 most probable trajectories from TEM$^{wt}$ → TEM* (Weinreich *et al*., 2006). It is important to highlight that—among all five mutations—the highest increase in cefotaxime resistance on TEM$^{wt}$ is conferred by the destabilizing G238S mutation (Weinreich *et al*., 2006). Thus, the nature of mutations (stabilizing or destabilizing) appears pivotal—from the standpoint of the least action principle—in assessing the efficiency of protein evolvability. Nevertheless, not all mutation can be destabilizing. In fact, the equilibrium between stabilizing and destabilizing mutations is ruled—at the molecular level—by the underlying physical principles that guide protein evolution, specifically the thermodynamic hypothesis fulfillment (Vila, 2025), which enabled us to set up a threshold for the allowed free energy changes upon mutations ($|\Delta\Delta G|$ ~7.4 kcal/mol)—beyond which a protein will unfold or become nonfunctional (Martin & Vila, 2020). On the other hand—at the biological level—such equilibrium is controlled by factors such as population dynamics (Poelwijk *et al*., 2007), protein-protein interactions (Jubb *et al*., 2017), the biophysical and genetic environment (Buda *et al*., 2023b; Fuentes-Ugarte *et al*., 2025), the way mutations and selection pressures interact across time (Di Bari *et al*., 2024), environmental changes (Dishman & Volkman, 2018; Vila, 2020), *etc*.

An examination of Eq. (1) reveals that the functional ($\Gamma_j^\mu$) is not invariant to the order of mutations. As a result, the implementation of the 'least action principle' to address evolutionary challenges—such as for β-lactamase—will be severely restricted, as knowledge of all $\Delta\Delta G_k^\mu$ values for each potential trajectory is required. This limitation illustrates the complexity of evolutionary processes, where multiple components must be considered simultaneously to be resolved. The fact that Weinreich *et al*. (2006) have limited their analysis to merely five mutations, despite acknowledging the potential for additional ones (DePristo *et al*., 2005), only exacerbates the problem of finding an accurate solution. Despite all the aforementioned drawbacks, the 'least action principle' could be highly useful in identifying trajectories that exhibit the highest level of evolutionary efficiency in a variety of contexts. This could be the case of the evolution of proteins from *m*-ancestors (with $m \geq 1$) to the same (see Figure 2**b**) or similar target sequence—a process that could be referred to as 'parallel' or 'convergent' depending on the ancestor origin (Bolnick *et al*., 2018; Cerca, 2023).

Overall, wild-type proteins that exhibit greater stability ($\Delta G_{wt}$) are more likely to evolve 'efficiently' towards a new function, as they can accommodate as many destabilizing mutations as possible—according to the principle of least action. In other words, the principle of least action imposes limits on the likelihood of a possible target sequence being reached efficiently and, hence, on protein evolvability.

### *The role of robustness*

The analysis above presents an alternative perspective on the influence of robustness on protein evolvability (Wagner, 2005b; Lenski *et al*., 2008; Wagner, 2008; Bloom & Arnold, 2009; Arnold, 2009; Masel & Trotter, 2010; Mayer & Hansen, 2017). For example, our results suggest that robustness is a *sine qua non* condition for the validity of the least action principle, which holds that protein evolution needs to be as efficient as possible. This efficiency calls for the emergence of destabilizing mutations that will speed up the folding rate (Vila, 2023b). The analysis of the β-lactamase gene evolution, TEM$^{wt}$ → TEM* (Wang *et al*., 2002; Weinreich *et al*., 2006), is an example of the latter, although it is not unique. In fact, the same pattern of substitutions appears to be present in other enzymes—including HIV protease (Condra *et al*., 1995) and DNA gyrase B protein (Blance *et al*., 2000)—that gain function at the expense of decreasing their thermodynamic stability. Robustness will become an important protein trait in evolution because it will enable both the emergence of destabilizing mutations and the fulfillment of the least action principle. Therefore, robustness will facilitate the emergence of novel evolutionary paths efficiently.

At this point, it is necessary to draw attention to a limitation of the analysis. Our methodology does not differentiate between neutral or harmful destabilizing mutations, *i.e.*, whether they are isolated or associated, respectively, with phenotypes that matter for fitness. As a result, we cannot answer fundamental evolutionary questions regarding the significance of neutral versus harmful mutations for protein evolvability and the role of robustness in this context (Lenski *et al*., 2008). Unfortunately, the analysis of these topics is beyond our current scope.

On the whole, the results of our analysis fully agree with pieces of evidence that protein stability promotes evolution (Bloom *et al*., 2006; Tokuriki & Tawfik, 2009; Bloom & Arnold, 2009; Arnold, 2009; Romero & Arnold, 2009; Ota *et al*., 2018). Regarding this extensive agreement with previous perspectives, Richard Feynman's (1948) comment (pp 367), "*There are, therefore, no fundamentally new results. However, there is a pleasure in recognizing old things from a new point of view*," is especially important to be highlighted here.

### *The role of epistasis*

It is usually claimed that epistasis may impose a specific mutational order, and the evolutionary paths connecting two natural β-lactamase variants are cited as an example of this process (Lipsh-Sokolik & Fleishman, 2024). In this case, how do epistasis effects add to the principle of least action? The following is the answer to such a vital question: epistasis governs the order in which mutations occur in possible evolutionary paths, whereas the principle of least action enables us to determine which of all conceivable trajectories is most likely to evolve efficiently. The impact of this conjecture—about the role of the epistasis in determining the most likely evolutionary path—results from our assumption about the distribution of Boltzmann factors, also known as relative probabilities ($P$), along any evolutionary path ($\mu$). For example, the sequence of Boltzmann factors for an arbitrary evolutionary path to go from TEM$^{wt}$ to TEM* ($P_{TEM^{wt} \rightarrow TEM^*}$) upon 5 mutations is given by:

$$P_{TEM^{wt} \rightarrow TEM^*} = \prod_{x \neq y}^{5} P_{xy} \qquad (2)$$

where $P_{xy} \sim e^{\beta \Delta\Delta G_{xy}}$, with $\Delta\Delta G_{xy} = (\Delta G_y - \Delta G_x)$ representing the Gibbs free energy change of the protein marginal stability after a single-point mutation (Vila, 2022). In this equation, $P_{xy}$ denotes the relative probability of obtaining protein $y$ subsequent to a single-point mutation in protein $x$. Consequently, these proteins could differ not only in the identity of a given residue of their sequence but, more significantly, of their structure and function due to the appearance of epistatic interactions. The latter suggests that protein evolution should follow a Markovian process—a time-independent stochastic process in which the probability of each step relies exclusively on the preceding step. It should be noted that Equation (2)—which we conjecture obeys a Markovian process—is independent of the elapsed time between mutations. Then, what can we say if the elapsed time between mutations takes not just minutes or hours but generations? Even in this scenario, the order of mutations during protein evolution should remain important because each non-silent single-point mutation will change the protein marginal stability at each step, influencing its likelihood of accommodating additional mutations or, as a result, adopting new functions throughout evolution. In other words, the order of mutations will impact the evolvability of proteins regardless of the elapsed time between them.

We should be aware of the fact that the above assumption—that Eq. (2) obeys a Markovian process—is one among other possible distributions of relative probabilities ($P$). For example, according to Weinreich *et al.* (2006) Equation (2) follows a probability distribution that is statistically independent of all previous mutations. There is a solid reason for these authors to adopt this conjecture. Indeed, if the modeling of protein sites' evolution depends on knowing all of the other sites in the sequence—as occurs when the epistatic interactions are taken into account—then their mathematical solution becomes a daunting task (Ashenberg *et al.*, 2013). What are the main drawbacks of applying Weinreich *et al.*'s (2006) conjecture to avoid such a problem? We will briefly mention two: Firstly, it ignores the importance of epistasis throughout evolution (Nasrallah *et al.*, 2011; Johnson *et al.*, 2023). Secondly, it assumes that the order of mutations is inconsequential at each stage of the evolutionary process, despite evidence on the contrary. Indeed, proteins may unfold or become nonfunctional if the order of the mutation is changed (Tokuriki & Tawfik, 2009; Domingo *et al*., 2019; Buda *et al*., 2023a; Buda *et al*., 2023b; Vila, 2025). Thus, for Eq. (2), all evidence indicates a Markovian process rather than a statistically independent model of all earlier mutations (Weinreich *et al*., 2006) should provide a better description of the protein evolvability process. This conclusion should contribute to the existing debate over whether protein sequence evolution is or is not a Markovian process (Kosiol & Goldman, 2011; Rizzato *et al*., 2016), a topic that is beyond the scope of this paper.

All things considered, we can conclude that (at the molecular level) every evolutionary mutational path, as specified in Eq. 2, follows—due to the epistatic interactions—a Markovian process. If each mutation is statistically independent of all previous ones, *i.e.*, not a Markovian process, then the functional ($\Gamma_j^\mu$) would become invariant to the order in which mutations occur, according to an examination of Eq. (1). Under this scenario, $\Gamma_j^\mu$ transforms into $\Gamma_j$; consequently, utilizing the principle of least action to determine the most 'efficient' evolutionary trajectory ($\mu$) for specific convergent evolutionary processes—such as β-lactamase—would be futile, as all trajectories would be equivalent, despite evidence suggesting otherwise: nature adheres to preferential pathways (Feynman *et al*., 1963; Stávek *et al*., 2002; Karl, 2012).

**Conclusions**

Our analysis indicates that the appearance of destabilizing mutations happens to be a necessary condition for proteins to evolve efficiently. We arrive at this conclusion after noticing that the presence of destabilizing mutations not only speeds the evolutionary process but also offers a rationale for the role of the principle of least action in enabling us to detect trajectories that will evolve efficiently. As a necessary condition for all of this to happen, protein robustness arises as a critical trait; otherwise, proteins will unfold or become nonfunctional in the presence of destabilizing mutations. In addition to everything said above, we proved how important epistasis is in establishing the mutation order, which plays a critical role in the determination of the most 'efficient' trajectory. On the whole, we can conjecture, without losing generality, that the main consequence of 'epistasis' and the 'principle of least action'—together—is to limit the possible evolutionary paths and trajectories as well as enable us to identify trajectories leading to novel protein sequences characterized by possessing a higher evolutionary efficiency rather than greater stability or fewer mutations. This implies that protein evolution may be more predictable or reproducible than previously thought. This insight could reshape our understanding of evolutionary biology at the molecular scale and also provide a rationale for how Darwinian selection might regard the numerous different mutational trajectories available. However, it is crucial to highlight that the model we presented is merely one of many that could explain protein evolution at the molecular level. Hence, the universality of our conclusions, along with their potential failure as the evolutionary system—and their surroundings—become more complex, will be ascertained by forthcoming studies.

## Acknowledgement

The author acknowledges support from the Institute of Applied Mathematics San Luis (IMASL), the National University of San Luis (UNSL), and the National Research Council of Argentina (CONICET).

## Founding

This research did not receive any specific grant from funding agencies in the public, commercial, or not-for-profit sectors.

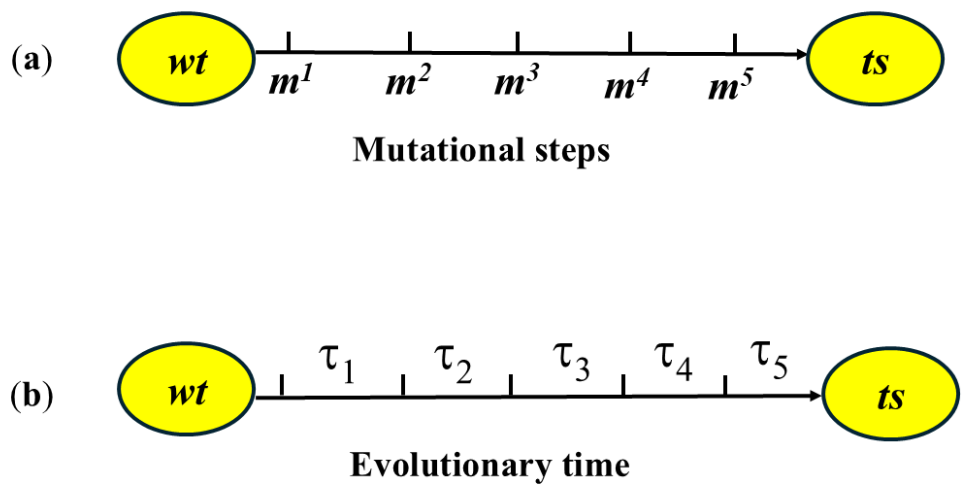

**Figure 1**. This figure shows, as a cartoon, how a wild-type protein sequence (*wt*) evolves into a target sequence (*ts*) after a series of five consecutive mutations ($m^x$, with $x$ = 1 to 5). This process defines the evolutionary 'path' displayed in panel (**a**). Panel (**b**) depicts a 'trajectory'—based on the evolutionary path defined in panel (**a**)—which is defined in terms of the folding rate ($\tau_x$) after each mutational step ($m^x$); see Eq.(1) for further details.

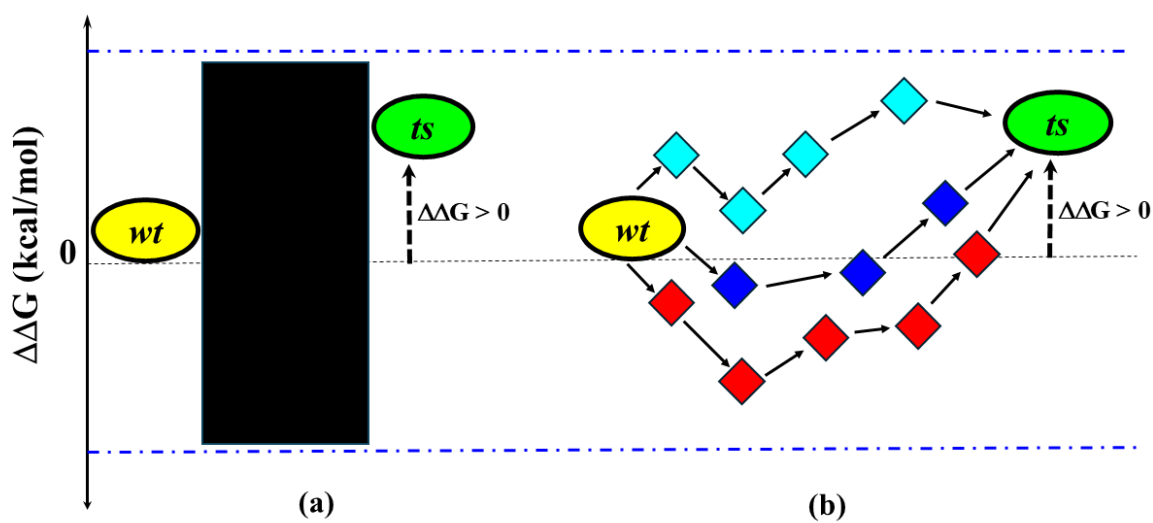

**Figure 2**. An illustration of two possible approaches to analyze protein evolution—from a wild-type (*wt*) to a target sequence (*ts*)—is displayed. In this figure, the horizontal blue dash-dot lines designate a universal threshold for $\Delta\Delta G$ (~ ±7.4 kcal/mol), beyond which any protein will unfold or become nonfunctional (Martin & Vila, 2020). From a thermodynamic viewpoint, the total free energy change for such a process ($\Delta\Delta G$) is a state function (Vila, 2024b), and hence, the only states that matter are the initial (*wt*) and the final one (*ts*), respectively (Vila, 2025); this strategy enables us to use the black box representation displayed in panel (**a**). Panel (**b**) depicts an alternative strategy in the form of a cartoon, with three possible trajectories—linking the *wt* with the *ts* sequence—highlighted by colored cyan, blue, and red rhombuses. Each of these evolutionary processes consists of 5, 4, and 6 mutational steps (indicated by black arrows), respectively. The main text addresses the advantages and disadvantages of each approach displayed in panels (**a**) and (**b**).